\documentclass{article} 
\usepackage{iclr2026_conference,times}

\usepackage{amsmath,amsfonts,bm}

\def\eqref#1{equation~\ref{#1}}

\def\1{\bm{1}}

\DeclareMathAlphabet{\mathsfit}{\encodingdefault}{\sfdefault}{m}{sl}
\SetMathAlphabet{\mathsfit}{bold}{\encodingdefault}{\sfdefault}{bx}{n}

\usepackage{hyperref}
\usepackage{url}

\title{Black Box Absorption:\\
LLMs Undermining Innovative Ideas}

\iclrfinalcopy

\author{Wenjun Cao \\
Independent Researcher \\
\texttt{wenjun.cao.research@gmail.com}
}

\begin{document}

\maketitle

\begin{abstract} Large Language Models are increasingly adopted as critical tools for accelerating innovation. This paper identifies and formalizes a systemic risk inherent in this paradigm: \textbf{Black Box Absorption}. We define this as the process by which the opaque internal architectures of LLM platforms, often operated by large-scale service providers, can internalize, generalize, and repurpose novel concepts contributed by users during interaction. This mechanism threatens to undermine the foundational principles of innovation economics by creating severe informational and structural asymmetries between individual creators and platform operators, thereby jeopardizing the long-term sustainability of the innovation ecosystem. To analyze this challenge, we introduce two core concepts: the idea unit, representing the transportable functional logic of an innovation, and idea safety, a multidimensional standard for its protection. This paper analyzes the mechanisms of absorption and proposes a concrete governance and engineering agenda to mitigate these risks, ensuring that creator contributions remain traceable, controllable, and equitable. \end{abstract}

\section{Introduction}
Large Language Models are rapidly becoming integral to modern productivity and creation. They have demonstrated substantial efficiency gains across diverse domains, including office automation, professional writing, and software development \citep{10.1093/qje/qjae044, doi:10.1126/science.adh2586, dellacqua2023navigating, chatterji2025how}. Extending beyond simple automation, they are increasingly used as collaborative partners in innovative tasks, from creative work and art generation to scientific research and discovery \citep{Anantrasirichai_2021, novikov2025alphaevolvecodingagentscientific, si2025can, Hubert2024, horton2023largelanguagemodelssimulated, anthis2025llmsocialsimulationspromising}. Individuals and organizations provide original ideas during interactive generation to refine concepts and accelerate the innovation lifecycle.

However, the new paradigm rests on an assumption that is often overlooked: the originality and privacy of ideas shared during these interactions are robustly protected.\citep{king2025userprivacylargelanguage, lukas_analyzing_2023, mireshghallah2024can, mireshghallah2025positionprivacyjustmemorization, ngong2025protectingusersthemselvessafeguarding, tran2025understandingprivacynormsllmbased, 10.1145/3706598.3713540, green2025leakythoughtslargereasoning, yi2025privacyreasoningambiguouscontexts, zhang2025searchingprivacyrisksllm, shao2024privacylens, google2024_generativeai_privacy, tamkin2024clioprivacypreservinginsightsrealworld, huang2025_values_in_the_wild} We argue that the foundational principles of innovation, namely the ability to secure and control novel concepts, are being undermined by the structure of the current ecosystem.\citep{bommasani2022opportunitiesrisksfoundationmodels, 10.1145/3442188.3445922, solaiman2023gradientgenerativeairelease, sastry2024computingpowergovernanceartificial, bommasani_considerations_2024, white2024modelopennessframeworkpromoting, Casper_2024, kapoor2024societalimpactopenfoundation, anderljung2023frontierairegulationmanaging}

The threat originates from the complex and opaque nature of LLM platforms. Protection of original ideas is not absolute. User inputs can be routed through internal systems for detection, annotation, utilization, and retraining \citep{han2025reinforcementlearninguserfeedback, liu2025userfeedbackhumanllmdialogues, anthropic2025_claude4_system_card, google2025_delivering_trusted_secure_ai}. Terms of Service and Privacy Policies frequently grant broad licenses to the provider \citep{oakley2005fairness}. Combined with the complexity of deep learning models, the overall platform behaves as a black box from the perspective of the user \citep{Burrell2016-BURHTM, 10.5555/2717112}. The user lacks verifiable insight into how ideas are processed, stored, or repurposed. The opacity is particularly consequential because the assets at risk, namely functional ideas and processes, often fall into a legal gap outside traditional copyright, which protects expression and not function \citep{pasquale2024consent, kyi2025governancegenerativeaicreative}. We refer to the combination of technical opacity and potential legal ambiguity as \textbf{Black Box Absorption}.

It is important to distinguish this topic from adjacent work on LLM privacy. Most research concentrates on training data leakage and privacy-preserving inference.\citep{anthropic2025confidential, google2025_delivering_trusted_secure_ai} Training data leakage, including memorization and regurgitation of sensitive information\citep{lukas_analyzing_2023, shokri2017membershipinferenceattacksmachine, yu2023bagtrickstrainingdata, shi2024detectingpretrainingdatalarge, carlini2023extractingtrainingdatadiffusion, carlini2021extractingtrainingdatalarge, kandpal2023largelanguagemodelsstruggle, kim2025detectingtrainingdatalarge}, does not directly address the live internal pipelines of deployed systems. The sources of training data are varied and often unrelated to interactive idea generation. Our analysis, therefore, focuses on the systemic risk of idea leakage within the deployment pipeline itself, where internal detection, annotation, and retraining workflows can operate as a continuous absorption mechanism.

The societal impact is clear in the economics of innovation \citep{10.1287/isre.1070.0138, PARTHA1994487, heller1998can, 10.1016/S0278-0984(02)11030-3, 10.1257/jep.5.1.29, TEECE1986285, Arrow1972, 10.1093/icc/dtq023}. Black Box Absorption creates a setting in which the platform becomes the central locus of value capture by drawing on unprotected ideas from many users. To analyze and safeguard this process, we introduce the idea unit as the specific asset at risk and propose idea safety as a multidimensional standard to protect such assets once they are articulated. Idea safety rests on three verifiable principles: traceability, control, and equitability.

Building on this perspective, the paper makes three contributions. It identifies and formalizes the systemic risk of Black Box Absorption that arises when deployment pipelines internalize valuable creator contributions. It develops the idea unit as a unit of analysis defined by a functional effect that specifies what is exposed during interaction and how it can diffuse. It advances idea safety as a deployable standard grounded in traceability, control, and equitability. This standard is used to trace the lifecycle of idea units on current platforms, analyze the economic consequences, and articulate a governance agenda that supports sustained innovation.

\section{Economic and Conceptual Foundations}

\subsection{Economics of Innovation Knowledge}

Large Language Models increasingly act as engines of innovation\citep{10.1093/qje/qjae044}. They can lower search costs, accelerate recombination of prior knowledge, and shorten the path from concept to workable draft \citep{bommasani2022opportunitiesrisksfoundationmodels}.

Innovation economics also explains why acceleration can coexist with new risks \cite {Arrow1972, 10.1257/jep.5.1.29, TEECE1986285}. Knowledge, once articulated, differs from ordinary goods. It is nonrival in use, inexpensive to copy, and easy to transmit across organizations and markets \citep{10.1257/jep.5.1.29, PARTHA1994487}. These properties enable diffusion that benefits society while reducing private value capture. To obtain support, an innovator must disclose enough detail for others to assess feasibility and impact, and the same disclosure can ease imitation or independent development \citep{Arrow1972, 10.1093/icc/dtq023, PARTHA1994487}. Legal protections mitigate some hazards but are incomplete \citep{heller1998can}. Many valuable elements of a concept, including methods, processes, problem framings, and strategic plans, are not fully covered, and legal protection often arrives slower than diffusion \citep{pasquale2024consent, kyi2025governancegenerativeaicreative}. Returns therefore depend on access to complementary assets such as data, compute, specialized labor, and distribution channels \citep{TEECE1986285}. Actors that control these assets can realize value faster and on a larger scale than originators who lack them \citep{narechania_machine_2021, narechania2024antimonopoly}.

Placing this framework in the context of LLM-mediated creation helps clarify the mechanism. Interactive use requires turning implicit know-how into explicit prompts, sketches, specifications, and critiques that the system can process \citep{king2025userprivacylargelanguage, tran2025understandingprivacynormsllmbased, Lucas2014ItsOA, 10.1093/iwc/iwae016, 10.1145/3613904.3642385}. Once expressed, similar economic properties often apply: replication is cheap, downstream reuse is feasible, and the originator’s control depends on institutional and technical safeguards that are often outside their reach \citep{Arrow1972, 10.1257/jep.5.1.29}. Platforms that host these interactions typically hold the complementary assets needed to transform promising content into deployable capabilities \citep{TEECE1986285, narechania_machine_2021}.

To analyze the dynamics with precision, we introduce the notion of an \emph{idea unit}: a minimally actionable piece of innovative content that, once articulated during interaction with a model, becomes subject to the forces described above. This unit serves as the basic object for reasoning about diffusion, control, and value realization throughout the rest of the paper.

\subsection{The Idea Unit}

\emph{An idea unit is identified by its functional effect, the transportable logic of a heuristic or reasoning pattern, rather than by its surface form as text, code, or a diagram}. It is the action-enabling structure that makes a proposal work: the organizing steps, the decision rule, the constraint that aligns choices, or the framing that unlocks a solution. Because it is defined by effect, the same unit can be rephrased many ways without altering what it enables.

Once a creator articulates reasoning to an LLM, the system receives a functional specification that can be separated from its wording and used in other contexts. The value and the vulnerability are both in that specification. It is easy to store, route, and recombine, and it travels across teams, products, and training pipelines. Control over the surface form does not guarantee control over the unit if its functional core has been captured \citep{usc17-102b, baker1879}.

Operationally, the idea unit is the object we track. It is the minimal slice of innovative content that, when exposed during interaction, can be evaluated, logged, sampled for review, curated into datasets, and generalized by models. Focusing on this unit allows clear reasoning about diffusion, control, and value realization without distraction from incidental phrasing.

\subsection{Idea Safety}

\emph{A system is idea safe when the creator can verify what happened to each unit, can govern its lifecycle after interaction, and can receive fair recognition if a unit contributes to improvement.} We propose a directional framework that turns this goal into concrete guarantees. The framework rests on three principles that map to the practical capabilities a creator must hold: agency over how each unit is handled, verifiable observability of its path through systems, and alignment of rewards when a contribution leads to improvement. The following paragraphs define these principles at an operational level and specify the conditions a platform must meet to satisfy them.

\paragraph{Control.}
Control requires that the creator retain agency over the idea unit after the interaction. Agency must be granular. A creator can decide, per unit, whether it may be retained for history, shared for safety review, used for training, or purged. Control includes the ability to correct misclassification, quarantine sensitive units, request redaction, and trigger unlearning or equivalent remedies when a unit was used against intent. Choices must be effective, visible, and reversible within clear time bounds\citep{10.1145/3706598.3713540, yi2025privacyreasoningambiguouscontexts, mireshghallah2025positionprivacyjustmemorization, shao2024privacylens}.

\paragraph{Traceability.}
Traceability requires that every step in the handling of an idea unit is knowable and auditable. A creator should be able to see when the unit was logged, who or what processed it, whether it entered review queues, whether any version was curated into datasets, and whether it influenced models or tools. Traceability is not a promise in policy language; it is a record that links the unit to its provenance and downstream use so that questions about exposure, reuse, and influence have concrete answers \citep{NEURIPS2024_7af60ccb, park2023trakattributingmodelbehavior, bae2022if}.

\paragraph{Equitability.}
Equitability aligns incentives when a creator contributes an idea unit. If a unit with proper consent improves a model, a product, or an operational workflow, there should be a transparent path to recognition and value-sharing. The mechanism can take different forms, including royalties, credits, access, or other compensating benefits, but the principle is constant: contribution is not a free resource \citep{10.1257/pandp.20181003, vipra2023marketconcentrationimplicationsfoundation, narechania2024antimonopoly}. Equitability balances platform scale with creator authorship and helps ensure that the gains from diffusion do not erase the originator’s claim to value.

\section{The Lifecycle of an Idea Unit on LLM Platforms}

An idea unit, defined by novelty and value, is a distinct asset. Its vulnerability arises from the environment in which it is handled. Foundational research has established the black box as a core concept for analyzing the technical opacity of algorithms \citep{Burrell2016-BURHTM} and the socio-economic opacity of data-driven systems \citep{10.5555/2717112, doi:10.1177/2053951720935146}. Building on this, we introduce \textbf{Black Box Absorption} to define the process that occurs within this environment: the systemic internalization of idea units by platform providers. This absorption is not a single event but a multistage process enabled by standard internal operations. We trace the chronological lifecycle of an idea unit from the moment it is submitted. This pathway is synthesized from the operational details provided in public model reports and system cards, which document the key stages of data handling\citep{workshop2023bloom176bparameteropenaccessmultilingual, touvron2023llama2openfoundation, openai2024gpt4technicalreport, anil2023palm2technicalreport, team_gemini_2025, anthropic2025_claude4_system_card, comanici2025gemini25pushingfrontier, geminiteam2025geminifamilyhighlycapable, Paleyes_2022, info16020087}.

\subsection{Data Governance and User Licensing}

The process begins before any idea unit is typed. The legal gateway is established by the Terms of Service and the Privacy Policy. Users commonly grant the platform a worldwide, nonexclusive, royalty-free license to use, copy, modify, and create derivative works of their content for operating, providing, and improving services. While essential for legitimate functions such as caching, a broad improvement clause can create opacity. It legally empowers the use of user-generated content, including valuable idea units, for many internal purposes. Users often have limited visibility into how this license is interpreted or operationalized, which may create information asymmetry \citep{oakley2005fairness, tang2025darkpatternsmeetgui, 10.1145/3613904.3642385, pasquale2024consent, lukas_analyzing_2023, mireshghallah2025positionprivacyjustmemorization}.

\subsection{Data Ingestion and Interaction Logging}

After acceptance of terms, submission of an idea unit triggers the technical lifecycle. Ingestion has two phases: real-time processing and durable storage.

\paragraph{Real-time Interaction Processing.} In many deployments, the system processes the interaction in real-time\citep{Paleyes_2022, info16020087}. Before reaching the generative model, the prompt that contains the idea unit may be scanned by a lightweight classification model acting as a prefilter for policy violations\citep{sheth2023causalityguideddisentanglementcrossplatform, markov2023holisticapproachundesiredcontent, hoover2025dynamic, anthropic2025_claude4_system_card, google2025_delivering_trusted_secure_ai}. The generated response may then be scanned by a second classifier that checks the output for harmful content\citep{sheth2023causalityguideddisentanglementcrossplatform, markov2023holisticapproachundesiredcontent, hoover2025dynamic, 10646809, anthropic2025_claude4_system_card, google2025_delivering_trusted_secure_ai, nghiem-daume-iii-2024-hatecot, 10.1145/3599696.3612895, roy-etal-2023-probing, kumar2024watchlanguageinvestigatingcontent, huang2025contentmoderationllmaccuracy, gao2025iwriteviolatescontent}.

\paragraph{Persistent Interaction Logging.}
After these multistep checks, the complete interaction tuple consisting of the user prompt that contains the idea unit, the model response, and any internal safety flags may be written to operational logs\citep{Paleyes_2022, info16020087} \citep{king2025userprivacylargelanguage, tran2025understandingprivacynormsllmbased, mireshghallah2024can, lukas_analyzing_2023, tamkin2024clioprivacypreservinginsightsrealworld, huang2025_values_in_the_wild}. The processing often occurs within milliseconds\citep{Paleyes_2022, info16020087}. For the user, the interaction appears complete. For the platform, the lifecycle of the logged unit can continue \citep{tamkin2024clioprivacypreservinginsightsrealworld}. 

\subsection{Data Sampling, Review, and Annotation}

Operational logs may be treated as an active dataset. Automated triage and sampling can select specific interactions for human review, routing them to queues based on business needs.

\paragraph{Automated Triage and Sampling.}
Automated processes may filter and sample raw logs, distributing interactions to specialized human-in-the-loop workstreams \citep{Paleyes_2022, info16020087}. Selected data can be sent to annotators for reinforcement learning from human feedback. Annotators rate, rewrite, or rank model responses. Reviewers may be exposed to the raw content of the idea unit. High value and novel units could be attractive candidates for review because they represent complex prompts that are useful for training more capable models\citep{liu2024makesgooddataalignment, lee2024rlaifvsrlhfscaling, Huang_2024, dong2024rlhf, han2025reinforcementlearninguserfeedback, liu2025userfeedbackhumanllmdialogues, huang2025_values_in_the_wild}

\paragraph{Safety and Audit Review.}
Interactions flagged by automated classifiers can be prioritized and routed to internal teams or safety contractors. Reviewers audit classifier decisions, handle edge cases, and provide corrections. These judgments may be used to retrain and improve automated safety filters\citep{Casper_2024, raji2020closingaiaccountabilitygap, google2025_delivering_trusted_secure_ai}.

\paragraph{Quality Assurance and Annotation.} A separate sampling pathway can target interactions based on other criteria, including negative user feedback or heuristics for high quality and novelty. These samples may be sent to annotators who provide labels useful for future training\citep{Paleyes_2022, NIPS2015_86df7dcf, doi:10.1073/pnas.2305016120} \citep{king2025userprivacylargelanguage, tran2025understandingprivacynormsllmbased}.

\subsection{Data Curation and Model Retraining}

This stage systematizes potential absorption. Valuable data identified earlier is prepared and consumed to update models.

\paragraph{Dataset Curation and Compiling.}
Human-labeled data is aggregated and combined with data selected by automated quality filters. The set is cleaned and deduplicated to create a curated dataset for a future training cycle \citep{NIPS2015_86df7dcf, 10.1145/3180155.3182515, 10.1145/2481244.2481247, Paleyes_2022, info16020087}.

\paragraph{Model Retraining and Generalization.}
The curated dataset may be used for fine-tuning or for building a new pre-training corpus. During retraining, parameters are updated, allowing patterns, concepts, and knowledge within the idea unit to be encoded and generalized. The novel concept within the unit could become non-exclusive if generalized into parameters. The generalized content may then influence responses to other users, including others in similar domains\citep{openai2024gpt4technicalreport, touvron2023llama2openfoundation, team_gemini_2025, anil2023palm2technicalreport, carlini2021extractingtrainingdatalarge, kandpal2023largelanguagemodelsstruggle, carlini2023extractingtrainingdatadiffusion, shokri2017membershipinferenceattacksmachine, kim2025detectingtrainingdatalarge, mireshghallah2025positionprivacyjustmemorization, lukas_analyzing_2023, pasquale2024consent, comanici2025gemini25pushingfrontier, geminiteam2025geminifamilyhighlycapable}
Across this multistage pipeline, from legal consent to automated generalization, transparency can be limited in practice. Creators may lack practical means to verify whether an idea unit was selected, how it was used, or whether removal was effective, and opt-out mechanisms may be limited or absent in some cases \citep{oakley2005fairness, tang2025darkpatternsmeetgui, Burrell2016-BURHTM, doi:10.1177/2053951720935146}.

\section{Consequences of Black Box Absorption}

The absorption pathway, combined with the economics of innovation and the definition of the idea unit, can create systemic risks that shape behavior across the ecosystem. Rapid diffusion, incomplete protection, and concentrated implementation capacity may pressure creators into choices that reduce their ability to capture value.

\subsection{Adoption Pressure in Competitive Settings}

In competitive environments where peers use LLMs to accelerate work, declining to use such tools may become unattractive. Creators recognize that using interactive tools can raise throughput and quality, while abstaining can lead to slower iteration, weaker outputs, and loss of opportunities. At the same time, using these tools requires articulating functional content that can be stored, routed, and learned from. Short-run costs of abstention can be salient, while the risk of absorption is opaque, probabilistic, and delayed. Given this asymmetry, actors may adopt tooling even when idea units could be incorporated into platform pipelines without verification or control. Adoption may be individually rational yet may collectively expose originators to systematic value loss \citep{10.1287/isre.1070.0138, PARTHA1994487, TEECE1986285}.

\subsection{Untraceable and Asymmetrical Control}

The central hazard is \emph{untraceability}. Once a functional pattern is generalized into a model, no straightforward audit trail ties a later capability to a specific contributing interaction. A creator may observe outputs, features, or practices that resemble prior submissions, yet the route from contribution to effect can remain hidden, which weakens any basis for recourse or negotiation\citep{NEURIPS2024_7af60ccb, park2023trakattributingmodelbehavior, bae2022if, Burrell2016-BURHTM}. Two forms of asymmetry follow. First, informational asymmetry: platforms can observe selection, review, curation, and training decisions end-to-end, while creators lack visibility into how their idea units are used. Second, structural asymmetry: platforms command complementary assets such as compute, data, engineering capacity, and distribution, which convert content into outcomes at speed and scale that originators typically cannot match\citep{narechania2024antimonopoly, narechania_machine_2021, vipra2023marketconcentrationimplicationsfoundation, Kleinberg_2021}.

\subsection{Asymmetrical Value Realization}

These asymmetries can channel returns away from originators and toward asset holders \citep{Kalluri2020}. When the functional core of an idea is easy to diffuse and difficult to exclude, value capture depends less on authorship and more on control of implementation bottlenecks\citep{TEECE1986285}. Potential absorption could intensify this pattern. Idea units supplied during interaction may be filtered, refined, and embedded into systems that only the platform can deploy widely. Features and capabilities can then appear to originate within the platform boundary. At the same time, the creators who supplied the enabling logic cannot readily demonstrate influence or claim a share of realized value. Over time, value could concentrate in institutions with the means of execution. The long-run cost is not only distributive; incentives to articulate high-value idea units may weaken, which harms dynamic efficiency \citep{10.1257/jep.5.1.29}.

\section{An Idea Safety Agenda}

The risks of asymmetry and untraceable absorption necessitate a new governance framework. We outline the deployable Idea Safety agenda, which provides the strategic direction to address these challenges. It translates our economic lens and the definition of the idea unit into foundational guidance for both engineering and policy. The goal is to replace ad hoc promises with verifiable guarantees, establishing a system that aligns with how idea units are created, managed, and converted into capabilities on contemporary platforms.

\subsection{The Control Principle}

Control begins with the premise that agency does not end at submission. An articulated idea unit should remain subject to the originator’s choices over retention, review exposure, training use, and purging. In practice, this requires an interaction mode with non-retention available as a default, clear disclosures about what is recorded and for how long, and a dashboard available after interaction where decisions can be made at the level of individual units \citep{google2024_generativeai_privacy}. Where a unit was misrouted or mislabeled, the creator must be able to correct that status. Where a unit was used against intent, there must be an effective remedy, such as redaction from logs and unlearning of downstream use. These controls must be auditable and enforced within stated windows so that the agency is operational rather than symbolic \citep{10.1145/3706598.3713540, yi2025privacyreasoningambiguouscontexts, mireshghallah2025positionprivacyjustmemorization, shao2024privacylens, google2025_delivering_trusted_secure_ai}.

\subsection{The Traceability Principle}

Traceability is an engineering commitment. Every idea unit that is selected for review or training carries its provenance forward, and the system can report what happened to it. An attribution first pipeline binds each copy, transformation, and decision to a consented source so that the path from submission to influence is reconstructable. On this substrate, the platform should answer two classes of questions without guesswork: where a unit went, including logging, sampling, curation, and evaluation, and how it mattered, including whether and where it contributed to model behavior or tool configurations. When removal is warranted, the exact provenance supports targeted unlearning and verification that subsequent artifacts no longer rely on the unit. Attribution, influence accounting, and unlearning are facets of a single system that makes the handling of idea units inspectable and, when necessary, reversible \citep{NEURIPS2024_7af60ccb, park2023trakattributingmodelbehavior, bae2022if, google2025_delivering_trusted_secure_ai}.

\subsection{The Equitability Principle}

Equitability addresses who benefits when idea units improve systems. Contractual guarantees today are often stronger for organizational accounts, while personal accounts are frequently treated as sources of training material by default. The remedy is a baseline that applies to everyone. The use of an idea unit for improvement must require explicit consent, accompanied by a clear account of permitted reuse and attribution. When a consented unit improves a model, product, or workflow, the platform should convert that contribution into tangible value such as credits, access, royalties, or other fair consideration, according to published terms that are auditable and applied symmetrically. Records that exist for traceability also serve as the ledger for contribution, ensuring that recognition and value-sharing reflect actual use \citep{10.1257/pandp.20181003, vipra2023marketconcentrationimplicationsfoundation, narechania2024antimonopoly}.

\section{Discussion and Related Work}

\paragraph{LLM Deployment.} Public system cards and engineering accounts outline a recurring deployment pattern in which user prompts are screened by lightweight classifiers, routed to large models, and logged together with safety metadata for later review and improvement\citep{Paleyes_2022, info16020087, openai2024gpt4technicalreport, touvron2023llama2openfoundation, team_gemini_2025, anil2023palm2technicalreport, anthropic2025_claude4_system_card, comanici2025gemini25pushingfrontier, geminiteam2025geminifamilyhighlycapable, google2025_delivering_trusted_secure_ai, tamkin2024clioprivacypreservinginsightsrealworld, huang2025_values_in_the_wild, nghiem-daume-iii-2024-hatecot, 10.1145/3599696.3612895, roy-etal-2023-probing, kumar2024watchlanguageinvestigatingcontent, huang2025contentmoderationllmaccuracy, gao2025iwriteviolatescontent}. Legal gateways such as Terms of Service and Privacy Policies typically authorize broad reuse of user content for service improvement, which establishes the contractual surface through which interactions can enter internal pipelines\citep{oakley2005fairness, tang2025darkpatternsmeetgui, 10.1145/3613904.3642385, google2024_generativeai_privacy}. This literature clarifies infrastructure and licensing, but it does not model the functional granularity of what is at stake for creators. Our account centers the \emph{idea unit} as the object that travels through these pipelines and provides a framework for what must be auditable and controllable once it has been articulated.

\paragraph{LLM Privacy.} Research on privacy has disproportionately targeted training data leakage and inference-time confidentiality; recent large-scale analyses of the field confirm that the vast majority of work overlooks threats from post-interaction data handling, platform governance, and retraining pipelines \citep{mireshghallah2025positionprivacyjustmemorization, brown2022doesmeanlanguagemodel}. This dominant focus is evident in a growing body of work that quantifies memorization, membership inference, and secret extraction risks in parametric models and fine-tuning regimes \citep{carlini2021extractingtrainingdatalarge, carlini2023extractingtrainingdatadiffusion, shokri2017membershipinferenceattacksmachine, kandpal2023largelanguagemodelsstruggle, lukas_analyzing_2023, carlini2023quantifyingmemorizationneurallanguage}. Another strand explores encrypted or privacy-preserving inference that hardens the runtime pathway at the cost of substantial latency and compute overheads \citep{NEURIPS2024_264a9b3c, staab2024memorizationviolatingprivacyinference, castro2024privacypreserving, zhang2025cipherprune, hao2022iron, anthropic2025confidential, google2025_delivering_trusted_secure_ai}. These advances secure data and interactions, yet they leave unresolved the critical gap of what happens when valuable reasoning is legitimately observed, labeled, curated, and then generalized within platform retraining. Our work complements privacy guarantees by specifying post-interaction rights over the functional content of ideas rather than only their textual surface.

\paragraph{RLHF.} Studies of human feedback pipelines describe how platforms select interactions, collect ratings, derive preference data, and use it for alignment and capability gains\citep{NEURIPS2022_b1efde53, bai2022constitutionalaiharmlessnessai, liu2024makesgooddataalignment, lee2024rlaifvsrlhfscaling, Huang_2024, dong2024rlhf, han2025reinforcementlearninguserfeedback, liu2025userfeedbackhumanllmdialogues, tamkin2024clioprivacypreservinginsightsrealworld, huang2025_values_in_the_wild}. Operations research highlights curation, quality control, and dataset construction as critical determinants of downstream behavior \citep{NIPS2015_86df7dcf, 10.1145/3180155.3182515, 10.1145/2481244.2481247, doi:10.1073/pnas.2305016120, nazabal2020dataengineeringdataanalytics}. This literature explains how content is transformed into a training signal but remains agnostic about the creator’s claims once their reasoning has been integrated. We make those claims explicit by requiring traceability at the unit level and remedies such as targeted unlearning when use conflicts with consent.

\paragraph{Innovation Economics.}
Economic analyses of knowledge production emphasize nonrivalry, diffusion, disclosure incentives, and the centrality of complementary assets in value capture \citep{Arrow1972, 10.1257/jep.5.1.29, PARTHA1994487, TEECE1986285, 10.1093/icc/dtq023}. Work on digital platforms and competition shows how control of infrastructure, data, and distribution shapes appropriation \citep{narechania_machine_2021, narechania2024antimonopoly}. We import this lens into LLM-mediated creation and make the exposure unit explicit: once a functional heuristic is articulated to a model, it becomes a transportable asset whose returns depend on provenance, consent, and the holder of complementary capabilities. The proposed notion of \emph{idea safety} translates classic appropriability concerns into deployable engineering and governance requirements.

\paragraph{Social Impact of LLMs.}
Surveys and systematizations document the expanding scope of LLM use, reported productivity gains, and emerging societal effects across research, industry, and culture \citep{bommasani2022opportunitiesrisksfoundationmodels, 10.1093/qje/qjae044, bommasani_considerations_2024, sastry2024computingpowergovernanceartificial, tamkin2024clioprivacypreservinginsightsrealworld, huang2025_values_in_the_wild}. Critical analyses warn that aggregate performance metrics can obscure distributional consequences and that capability scaling can reshape creative practice \citep{10.1145/3442188.3445922, solaiman2023gradientgenerativeairelease, white2024modelopennessframeworkpromoting, kapoor2024societalimpactopenfoundation, anderljung2023frontierairegulationmanaging}. Our contribution engages these debates by identifying a concrete failure mode called Black Box Absorption, where the gains from diffusion may accrue within platform boundaries without verifiable pathways for recognition or control for the originators of functional ideas.

\paragraph{AI Risks.} Foundational critiques of the algorithmic black box argue that complex, opaque, and inscrutable learning systems create structural information asymmetries that frustrate accountability\citep{Burrell2016-BURHTM, 10.5555/2717112, doi:10.1177/2053951720935146, Casper_2024}. Policy-oriented work examines monopolization risk, concentration of critical inputs, and market power that can convert aggregate improvement into centralized value capture \citep{narechania2024antimonopoly, vipra2023marketconcentrationimplicationsfoundation, Kleinberg_2021}. Our framework operationalizes these concerns at the unit of contribution: by making provenance, influence, and remedial action primary system properties, it becomes possible to mitigate opacity without halting capability progress, and to counter homogenization by aligning improvement with auditable sources.

The present paper makes three contributions. It shifts the unit of analysis from datasets or conversations to idea units defined by functional effect, which explains why value can be at risk even when surface text is protected. It treats absorption as a deployment lifecycle phenomenon that spans legal gateways, logging, sampling, curation, and retraining, rather than a property of pretraining data alone. It specifies a deployable standard for idea safety, integrating control, traceability, and equitability as requirements for credible collaboration between creators and platforms. This synthesis complements privacy techniques, extends RLHF operations with provenance obligations, and grounds economic concerns in a concrete engineering target.

\section{Conclusion}

This paper studies a structural risk, called Black Box Absorption, in which interactive use of Large Language Models could allow providers to internalize the functional content of creator contributions in ways that are opaque and difficult to contest. We defined the idea unit as the minimal actionable content exposed during interaction, traced a plausible lifecycle through governance, ingestion, review, curation, and retraining pipelines, and analyzed how informational and structural asymmetries could shift value capture away from originators. Building on this perspective, we proposed the concept of idea safety as a deployable standard grounded in control, traceability, and equitability. The agenda's importance is both economic and ethical, as sustained innovation depends on credible guarantees that align incentives for contribution. Future work should deliver verifiable provenance and influence accounting, practical unlearning at the level of individual units, and consent and value-sharing mechanisms that operate by default. It should also include reproducible audits and benchmarks for compliance, as well as policy and contractual frameworks that make these guarantees enforceable across providers.

\bibliography{main}
\bibliographystyle{iclr2026_conference}


\end{document}